\documentclass{PoS}

\pdfoutput=1
\usepackage{url}
\usepackage[sort&compress,square,comma]{natbib}

\newcommand{\pflux}{ cm$^{-2}$ s$^{-1}$}
\newcommand{\lsi}{LS~I~+61$^{\circ}$~303}
\newcommand{\gev}{\,GeV}
\newcommand{\tev}{\,TeV}

\title{VERITAS observations of exceptionally bright TeV flares from \lsi{}}

\ShortTitle{Bright TeV flares from \lsi{}}

\author{\speaker{A. O'Faol\'ain de Bhr\'oithe}~for the VERITAS collaboration\thanks{veritas.sao.arizona.edu}\\
        DESY, Platanenallee 6, 15738 Zeuthen, Germany\\
        E-mail: \email{anna.ofaolain.de.bhroithe@desy.de}}

\abstract{The very-high-energy (VHE; E > 100 GeV) gamma-ray experiment, VERITAS, detected exceptionally bright flares from the high-mass X-ray binary \lsi{} during the period October\,--\,December 2014. \lsi{} is a known VHE gamma-ray source, the flux from which varies strongly with the orbital period of $\sim26.5$ days. The maximum VHE flux is found around ap\-astron (orbital phase $\sim0.6$) at a level typically corresponding to 10\,--\,15\% of the Crab Nebula flux (>300 GeV). During these most recent observations, relatively short (day scale), bright TeV flares were observed from the source around apastron in two orbital cycles (October and November). Both cases exhibited peak fluxes above 25\% of the Crab Nebula flux (>300 GeV), making these the brightest VHE flares ever detected from this source. In the last orbital cycle observed (December), the source had returned to its historical level of activity. The results of these VERITAS observations from 2014 will be presented. }

\FullConference{The 34th International Cosmic Ray Conference,\\
		30 July- 6 August, 2015\\
		The Hague, The Netherlands}

\begin{document}
\section{Introduction}

The class of TeV-emitting high-mass X-ray binaries (HMXBs) consists of only a few sources: LS 5039 \cite{2005Sci...309..746A}, PSR B1259-63 \cite{2005A&A...442....1A}, \lsi{} \cite{Albert2006}, HESS J0632+057 \cite{2009ApJ...698L..94A}, and the newest member of the class 1FGL J1018.6-5856 \cite{2015A&A...577A.131H}. Only the compact object of PSR B1259-63 has been firmly identified as a pulsar, while the natures of the compact objects in the other systems have not yet been unambiguously determined. Consequently, the fundamental setup that produces the TeV emission along with its characteristic variability on the timescale of one orbital period remains unknown.

\lsi{} is the only known TeV binary in the Northern Hemisphere that has a sufficiently short orbital period to allow for regular study with TeV instruments. Located at a distance of $\sim2$\,kpc \cite{1991AJ....101.2126F}, the system comprises a B0 Ve star and a compact object \cite{HandC1981, Casares2005}. The observed emission is variable and modulated with a period of $P \approx 26.5$ days, believed to be associated with the orbital structure of the binary system \cite{Albert2006, Esposito2007, VERITASLSIDetection, Abdo2009, LiXray, 2015A&A...575L...9M}. Radial velocity measurements show the orbit to be elliptical $e = 0.537\pm0.034$, with periastron occurring around phase $\phi=0.275$ and apastron at $\phi=0.775$ \cite{Aragona2009}. The periastron distance between the star and the compact object is estimated at $2.84 \times 10^{10}$\,m (0.19\,AU) and the apastron distance at $9.57 \times 10^{10}$\,m (0.64\,AU) \cite{2013A&ARv..21...64D}. The inclination of the system is expected to lie in the range $10^\circ$\,--\,$60^\circ$ \cite{2013A&ARv..21...64D}, but the lack of exact knowledge of this parameter leads to considerable uncertainty of the other orbital parameters.

In this work, the results of the VERITAS campaign on \lsi{} in late 2014 are presented. During this time, VERITAS observed historically bright flares from the binary around apastron, with the source exhibiting flux levels a factor of 2\,--\,3 times higher than ever previously observed.

\section{Observations}
VERITAS is an imaging atmospheric-Cherenkov telescope (IACT) array located at the Fred Lawrence Whipple Observatory (FLWO) in southern Arizona (1.3\,km a.s.l., 31$^{\circ}$40'\,N, 110$^{\circ}$57'\,W). It consists of four 12\,m diameter Davies-Cotton design optical telescopes, each with a reflector composed of 345 tessellated hexagonal mirror facets that focus light onto a camera at the focal plane with a field of view of $3.5^\circ$. It is sensitive to photons with energies from 85\gev{} to $>30$\tev{} with an energy resolution of 15\,--\,25\%, an angular resolution of $<0.1^\circ$ at 1\tev{}, and the ability to detect a 1\% Crab Nebula source in approximately 25 hours\footnote{\url{http://veritas.sao.arizona.edu}} \cite{parkicrc}. For a full description of the hardware components and analysis methods utilized by VERITAS, see \cite{VERITAS, KiedaVTSUpgrade, VERITASLSIDetection}.

\begin{figure}[ht]
\centering

\includegraphics[width=0.8\textwidth]{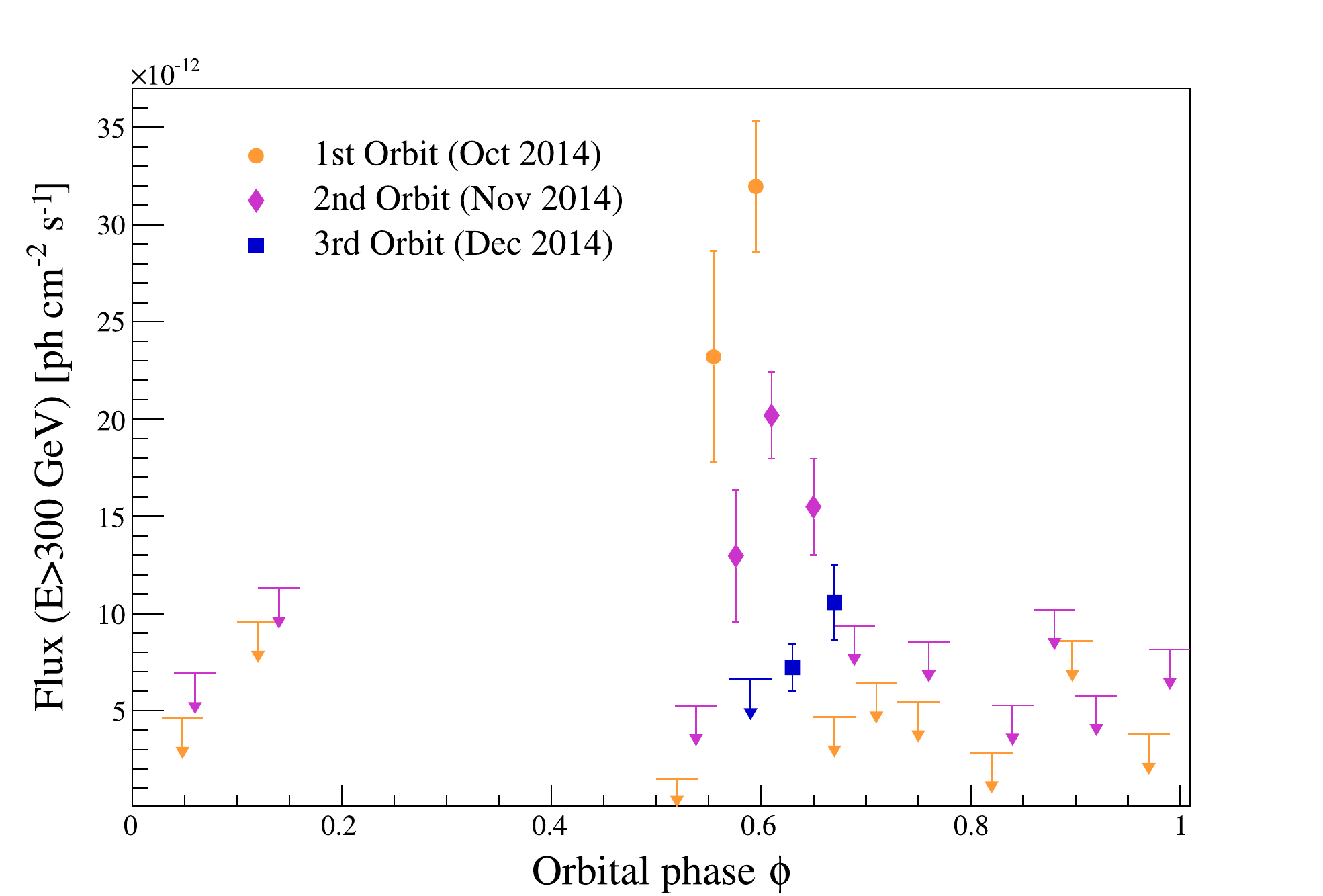}
\caption{Light curve of \lsi{} during the 2014 observation season in nightly bins. The data for the first orbit (October) are shown with orange circles, the data for the second orbit are shown with purple diamonds, and the data for the third orbit are shown with blue squares. Flux upper limits at the 99\% confidence level (using the unbounded approach of \cite{Rolke}) are shown for points with $<3\,\sigma$ significance and are represented by arrows.
}
\label{f:fluxphase}
\end{figure}

In the 2014 season, VERITAS observations of \lsi{} were taken from October 16 (MJD 56946) to  December 12 (MJD 57003) for a total of 24.7 hours of quality selected livetime. These observations covered three separate orbital periods, sampling the orbital phase regions of $\phi = 0.5-0.2$ (see Figure~\ref{f:fluxphase} and Table~\ref{t:fluxphase}). Over the entire set of observations, a total of 449 excess events above an energy threshold of 300\gev{} were detected above background, equivalent to a statistical significance of 21 standard deviations above background ($21\sigma$, calculated using Equation 17 of \cite{LiMa}).

\begin{table}
\centering
\begin{tabular}{ccc} \hline
Date observed [MJD] & Orbital phase & Flux($>300$\gev{}) [$\times 10^{-11}$ cm$^{-2}$ s$^{-1}$] \\ \hline
56946 & 0.52 & $<$0.15\phantom{ $\pm$ 0.00}\\  
56947 & 0.55 & \phantom{$<$}2.32 $\pm$ 0.54  \\
56948 & 0.60 & \phantom{$<$}3.20 $\pm$ 0.34 \\
56950 & 0.67 & $<$0.47\phantom{ $\pm$ 0.00} \\
56951 & 0.71 & $<$0.64\phantom{ $\pm$ 0.00} \\
56952 & 0.75 & $<$0.55\phantom{ $\pm$ 0.00} \\
56954 & 0.82 & $<$0.28\phantom{ $\pm$ 0.00} \\
56956 & 0.90 & $<$0.86\phantom{ $\pm$ 0.00} \\
56958 & 0.97 & $<$0.38\phantom{ $\pm$ 0.00} \\
56960 & 0.05 & $<$0.46\phantom{ $\pm$ 0.00} \\
56962 & 0.12 & $<$0.96\phantom{ $\pm$ 0.00} \\
56973 & 0.54 & $<$0.53\phantom{ $\pm$ 0.00} \\
56974 & 0.58 & \phantom{$<$}1.30 $\pm$ 0.34 \\
56975 & 0.61 & \phantom{$<$}2.02 $\pm$ 0.22 \\
56976 & 0.65 & \phantom{$<$}1.55 $\pm$ 0.25 \\
56977 & 0.69 & $<$0.94\phantom{ $\pm$ 0.00} \\
56979 & 0.76 & $<$0.85\phantom{ $\pm$ 0.00} \\
56981 & 0.84 & $<$0.53\phantom{ $\pm$ 0.00} \\
56982 & 0.88 & $<$1.02\phantom{ $\pm$ 0.00} \\
56983 & 0.92 & $<$0.58\phantom{ $\pm$ 0.00} \\
56985 & 0.99 & $<$0.82\phantom{ $\pm$ 0.00} \\
56987 & 0.06 & $<$0.69\phantom{ $\pm$ 0.00} \\
56989 & 0.14 & $<$1.13\phantom{ $\pm$ 0.00} \\
57001 & 0.59 & $<$0.66\phantom{ $\pm$ 0.00} \\
57002 & 0.63 & \phantom{$<$}0.72 $\pm$ 0.12 \\
57003 & 0.67 & \phantom{$<$}1.06 $\pm$ 0.20 \\ \hline
\end{tabular}
\caption{VERITAS observations of \lsi{} in 2014}
\label{t:fluxphase}
\end{table}

During the first orbit observed (in October), the source presented the largest of its flares (referred to as ``F1'' from here on), beginning on 2014 October 17 (MJD 56947, $\phi = 0.55$) with emission reaching a peak of $(31.9 \pm 3.4_{\mathrm{stat}}) \times10^{-12}$\pflux{} on October 18 (MJD 56948). This is the largest flux ever detected from this source. Observations following the peak were limited by poor weather conditions and only recommenced on October 20 (MJD 56950), by which time the flux from the source had already decreased. During the second orbital passage in November, VERITAS detected another period of elevated flux (``F2'') from the source at similar orbital phases ($\phi = 0.5-0.6$) with peak emission of $(20.2 \pm 2.2_{\mathrm{stat}}) \times10^{-12}$\pflux{} on November 14 (MJD 56975).

The rise and fall times of the flares were determined by fitting Equation 7 of \citet{2010ApJ...722..520A} to the light curve of each orbit and fixing $t_0$ to the MJD of the observed peak. The fit to the first orbit is not very good, with a fit probability of $3 \times 10^{-7}$. Nevertheless, the rise and fall times of F1 are found to be $0.39 \pm 0.07$ days and $0.37 \pm 0.23$ days, respectively. The second orbit is well fit by this function, with a fit probability of $1 \times 10^{-1}$, possibly due to the better data sampling throughout this flare. The rise and fall times of F2 are thus found to be $0.65 \pm 0.13$ days and $0.83 \pm 0.12$ days, respectively. A piecewise defined exponential function was also fit to F1 and F2, but resulted in a very poor fit probability for both flares. Variability on a nightly timescale was tested using the method described in \cite{2013ApJ...779...88A}. Similar to their findings for this source, a hint of nightly variability at a significance level of $\sim3\,\sigma$ post trials was found in F1.

Follow-up observations conducted by VERITAS during the next month (2014 December 10\,--\,12) covered the orbital phases of $\phi=0.59-0.67$ and detected the source at a lower flux level, reaching only $(7.2 \pm 1.2_{\mathrm{stat}}) \times10^{-12}$\pflux{} around the orbital phase at which the flares were detected in the previous orbits. The observations during this month appear to exclude the type of peaked flaring behavior seen at the same phase range in the previous two orbital cycles, perhaps indicating some orbit-to-orbit variations in the system.

\begin{figure}[t]
\centering
\includegraphics[width=0.65\textwidth]{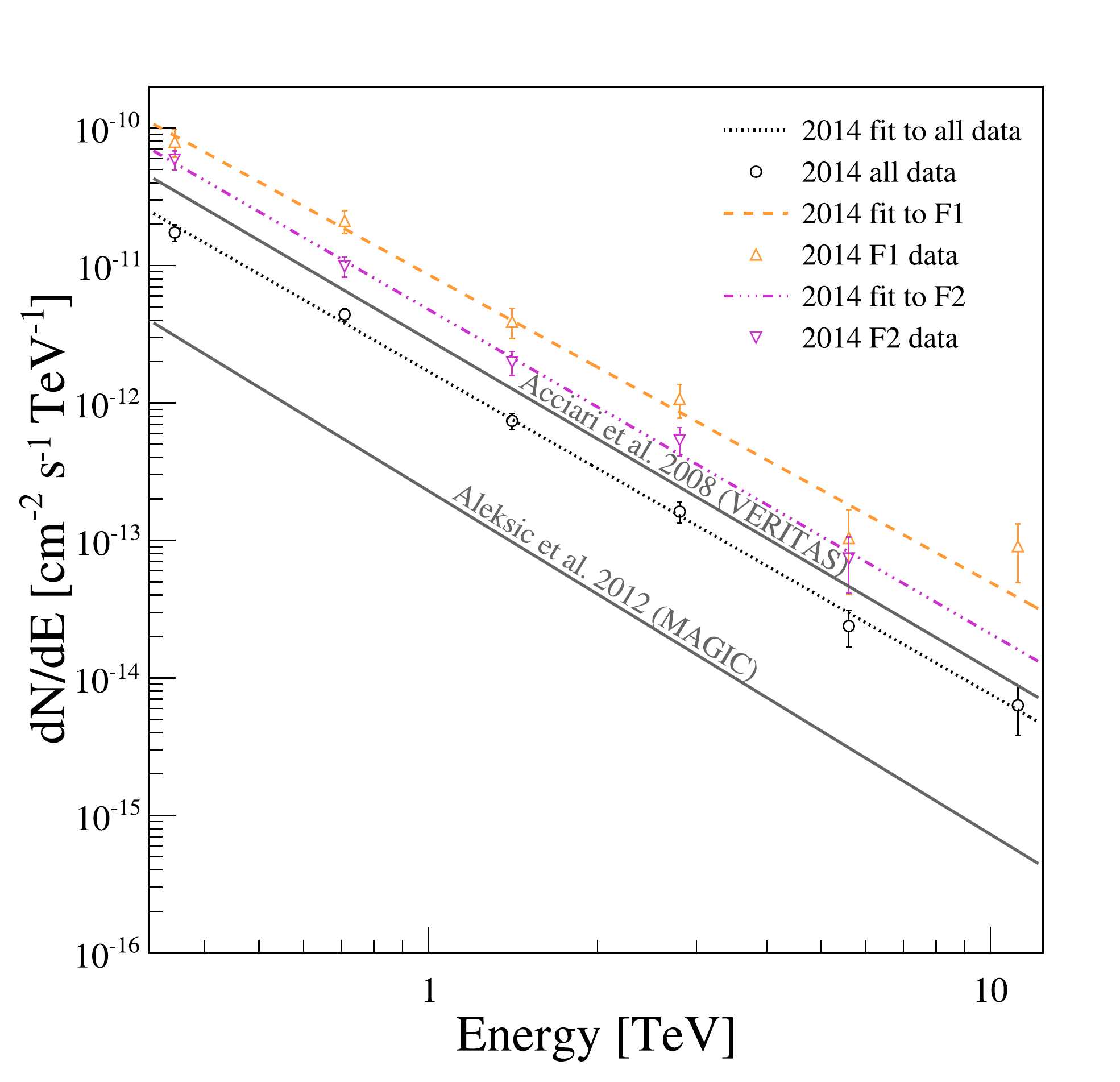}
\caption{Average and flare differential energy spectra of \lsi{} from the VERITAS 2014 observations, shown in comparison with the average spectra from \cite{VERITASLSIDetection} and \cite{Aleksic}.}
\label{spec}
\end{figure}

The average differential energy spectrum from all observations of \lsi{} during the 2014 observing season was found to be well fit with a power law of the form
\begin{equation}
\frac{dN}{dE} = N_0 \left( \frac{E}{1\mbox{\tev{}}} \right)^{\Gamma},
\end{equation}
in which $N_0$ is the normalization at the pivot energy of 1\tev{}, and $\Gamma$ is the spectral index. The measured parameters are consistent with past observations. Differential energy spectra were also extracted from F1 (October 17\,--\,18) and F2 (November 13\,--\,15) and show a similar spectral shape, albeit with a higher normalization constant. The parameters from the spectral fits are given in Table~\ref{t:specfits}. An uncertainty on the energy scale of 15\,--\,25\% results in a systematic uncertainty of $\sim50\%$ on the flux normalization and $\sim40\%$ on the integral flux, assuming a spectral index of -2.5. The systematic uncertainty on the spectral index is estimated at $\sim 0.3$, accounting for uncertainties on the collection efficiency, sky brightness, analysis cuts and simulation model. All spectra are shown in Figure~\ref{spec} along with previous spectral measurements for comparison.

\begin{table}[tb]
\centering
\begin{tabular}{ccc} \hline
Observations & Normalization [$\times 10^{-12}$ cm$^{-2}$ s$^{-1}$ TeV$^{-1}$] & Spectral index \\ \hline
All (average) & $1.7 \pm 0.7_{\mathrm{stat}} \pm 0.9_{\mathrm{sys}}$ & $-2.35 \pm 0.32_{\mathrm{stat}} \pm 0.3_{\mathrm{sys}}$ \\
F1 (Oct 17\,--\,18) & $8.6 \pm 1.0_{\mathrm{stat}} \pm 4.3_{\mathrm{sys}}$ & $-2.24 \pm 0.12_{\mathrm{stat}} \pm 0.3_{\mathrm{sys}}$ \\
F2 (Nov 13\,--\,15) & $4.8 \pm 0.4_{\mathrm{stat}} \pm 2.4_{\mathrm{sys}}$ & $-2.36 \pm 0.12_{\mathrm{stat}} \pm 0.3_{\mathrm{sys}}$ \\ \hline
\end{tabular}
\caption{Spectral parameters of the power law fits to the observations of \lsi{} in the energy range 0.3\,--\,20\tev{}.}
\label{t:specfits}
\end{table}

After the detection of F2 , VERITAS released an ATel \cite{2015VTSATEL} notifying the astronomical community of the historic flux levels and triggering observations by multiwavelength partners, as well as additional observations with the MAGIC TeV experiment. The data from this campaign are under analysis and will be presented in an upcoming publication. Initial results of TeV, GeV and X-ray observations during this campaign are presented in \cite{Karicrc}.

\section{Discussion and Conclusion}
The nature of the compact object in \lsi{} is not firmly established, so proposed models of the system cover a range of possibilities. In general, the models fall into one of two main categories: a microquasar ($\mu$Q) scenario or a pulsar binary (PB) scenario. The majority of both model types employ a leptonic origin of the observed non-thermal emission.

In the $\mu$Q scenario, non-thermal particle acceleration processes occur in the jet of an accreting compact object. Soft photons from the optical star and its accretion disk, and also from the synchrotron radiation in the jet can be upscattered by the ultrarelativistic electrons in the jet \cite{1538-4357-650-2-L123,Bednarek01102006}. Support for this type of model comes from the detection of an elongation of the source on the scale of tens of milliarcseconds in radio wavelengths \cite{Massi2001,Massi2004}, interpreted as a Doppler-boosted jet with precession \cite{Massi2013,2015A&A...575L...9M}.

The PB scenario utilizes the presence of a shocked wind in which particle acceleration is the result of the interaction between the stellar and the pulsar winds. For example, the author of \cite{Dubus2006} proposes the compact object to be a young pulsar with a spin-down rate of the order of $10^{36}$~erg s$^{-1}$ that generates a pulsar wind composed of principally monoenergetic electrons/positrons. The pulsar wind expands and shocks as it interacts with the stellar wind from the optical star. This shock has a ``bow'' or ``comet'' shape with a tail extending away from the optical star. The stellar photons can be upscattered to TeV energies by the particles accelerated at the shock. Support for this type of model comes from the detection of cometary-shaped radio emission, pointed away from the high-mass star \cite{Dhawan2006}.

A general PB scenario is presented by \cite{Paredes-Fortuny2014} describing an inhomogeneous stellar wind in which the Be star disc is disrupted and fragments, and the resulting clumps of the disc fall into the shock region, pushing it closer to the pulsar. The reduction in size of the pulsar wind termination shock could allow for increased acceleration efficiency on the timescale of a few hours, depending on the size and density of the disc fragments. Such a scenario could qualitatively account for the exceptionally bright TeV flares and orbit-to-orbit variations seen in \lsi{}.

The detection of pulsed emission from \lsi{} would unambiguously identify the source as a pulsar binary. While no pulsations have been detected to date, it is also possible that the dense stellar environment of the source could hinder such a detection. Regardless, further observations of \lsi{} with TeV instruments are necessary to fully understand the varying TeV emission from the source.
\vspace{2ex}

\small{
This research is supported by grants from the U.S. Department of Energy Office of Science, the U.S. National Science Foundation and the Smithsonian Institution, and by NSERC in Canada. We acknowledge the excellent work of the technical support staff at the Fred Lawrence Whipple Observatory and at the collaborating institutions in the construction and operation of the instrument. The VERITAS Collaboration is grateful to Trevor Weekes for his seminal contributions and leadership in the field of VHE gamma-ray astrophysics, which made this study possible. A.\ O'FdB acknowledges support through the Young Investigators Program of the Helmholtz Association. This study of \lsi{} was made possible by the Cycle 7 Fermi Guest Investigator program, grant number NNH13ZDA001N.
}

\bibliography{ofdb_lsI}

\end{document}